\documentclass[12pt]{article}
\usepackage[top=3cm,right=3cm,left=2cm,bottom=3cm]{geometry}
\usepackage{amsmath}
\usepackage{amssymb}

\newcommand{\at}{allowed transformations}

\newcommand{\tr}{transformation}
\newcommand{\gbe}{generalized Burgers equation}

\newcommand{\gen}[1]{\partial_{#1}}
\newcommand{\curl}[1]{ \{#1\} }

\newtheorem{thm}{Theorem}

\numberwithin{equation}{section}
\begin{document}

\title{\bf \vspace*{-1.5in}
Infinite-dimensional symmetries of a two-dimensional generalized Burgers equation}

\author{F. G\"{u}ng\"{o}r  \\ \small
Department of Mathematics, Faculty of Science and Letters,
\\
\small Istanbul Technical University, 34469, Istanbul, Turkey
\thanks{e-mail: gungorf@itu.edu.tr}}
\maketitle

\abstract{The conditions for a generalized Burgers equation
which a priori  involves nine arbitrary
functions of one, or two variables to allow an infinite dimensional symmetry
algebra are determined.  Though this algebra can involve up to two arbitrary functions
of time, it does not allow a Virasoro algebra. This result confirms that  variable coefficient generalizations of a non-integrable equation should be expected to remain as such.}
\section{Introduction}

In Ref. \cite{Gungor02-2} this author and Winternitz examined the conditions under which the generalized KP equation
\begin{equation}\label{GKP}
\begin{split}
& (u_t+p(t)uu_{x}+q(t)u_{xxx})_{x}+\sigma(y,t)u_{yy}+a(y,t)u_{y} \\
&+ b(y,t)u_{xy}+c(y,t)u_{xx}+e(y,t)u_{x}+f(y,t)u+h(y,t)=0.
\end{split}
\end{equation}
has  an infinite-dimensional symmetry algebra. In particular, it was shown that the canonical form of \eqref{GKP} obtained for
\begin{equation}\label{SCoef}
p=q=1,\quad \sigma=\varepsilon=\pm 1, \quad e=h=0
\end{equation}
allows the Virasoro algebra as
a symmetry algebra if and only if the coefficients satisfy
\begin{equation}\label{IGKP}
a=f=0,\quad b=b(t),\quad c=c_0(t)+c_1(t)y.
\end{equation}
Under these conditions, the equation can be transformed by a point transformation to the standard KP one. In the situation where the symmetry algebra becomes Kac-Moody type, the equation is slightly more general and we have in addition to $a=f=0$
\begin{equation}\label{KMT}
  b(y,t)=b_1(t)y+b_0(t),\quad c(y,t)=c_2(t)y^2+c_1(t)y+c_0(t).
\end{equation}

The purpose of this article is similar, namely to study the symmetry
properties of a \gbe {}
\begin{equation}\label{1.1}
\begin{split}
& (u_t+p(t)uu_{x}+q(t)u_{xx})_{x}+\sigma(y,t)u_{yy}+a(y,t)u_{y} \\
&+ b(y,t)u_{xy}+c(y,t)u_{xx}+e(y,t)u_{x}+f(y,t)u+h(y,t)=0,
\end{split}
\end{equation}
where we assume that in some neighbourhood we have
\begin{equation}\label{1.2}
p(t)\ne 0,\quad q(t)\ne 0,\quad \sigma(y,t)\ne 0.
\end{equation}
The other functions in \eqref{1.1} are arbitrary. We intend to determine the cases when eq.
\eqref{1.1} has an infinite-dimensional symmetry group.  More important, we would like to look at the possibility of  whether it can have a Kac-Moody-Virasoro structure. The presence of a Virasoro algebra as a subalgebra may exhibit a strong indication of the integrability of the equation. Indeed, this was the case for the GKP equation \eqref{GKP}-\eqref{IGKP}.

Painleve analysis can be performed for variable coefficient partial differential equations to decide about integrability or partial integrability, but in spite of computer algebra packages developed for this aim, the computations involved for such equations usually turn up to be  unmanageably lengthy.

Even though the algebra has no structure of a Virasoro algebra which is a typical property of integrable equations in 2+1 dimensions, the existence of an
infinite-dimensional symmetry group makes it possible to use Lie
group theory to obtain large classes of solutions.

The author also showed that the  Lie symmetry algebra of the two-dimen\-sion\-al generalized Burgers equation
\cite{Gungor01-2, Gungor02-1}
\begin{equation}\label{main}
(u_t+u u_x-u_{xx})_x+\sigma(t) u_{yy}=0
\end{equation}
which is a special case of \eqref{1.1} in which
$$p=1, \quad q=-1, \quad \sigma(y,t)=\sigma(t),\quad a=b=c=e=f=h=0$$
has a non-Abelian Kac-Moody structure and for arbitrary $\sigma$ is realized by
\begin{equation}\label{V}
\hat{\mathbf{V}}=X(f)+Y(g),
\end{equation}
\begin{subequations}
\begin{eqnarray}\label{A}
& X(f)&=f(t)\gen x+\dot{f}(t)\gen u, \\
& Y(g)&=g(t)\gen y-\frac{\dot{g}(t)}{2\sigma(t)}y\gen x-
\frac{d}{dt}\Bigl(\frac{\dot{g}(t)}{2\sigma(t)}\Bigr)y\gen u,
\end{eqnarray}
\end{subequations}
where $f(t)$ and $g(t)$ are arbitrary smooth functions and the primes denote time
derivatives. The algebra extends for several special forms of $\sigma(t)$ (see \cite{Gungor01-2, Gungor02-1} for the details).

In Section 2 we introduce "\at{}" that take equations of the form
\eqref{1.1} into other equations of the same class. That is, they
may change the unspecified functions in eq. \eqref{1.1}, but not
introduce other terms, or dependence on other variables. The \at{}
are used to simplify eq. \eqref{1.1} and transform it into eq.
\eqref{canon} that we call the "canonical \gbe{}" (CGB
equation). In Section 3 we determine the general form of the
symmetry algebra of the CGB equation and obtain the determining
equations for the symmetries. In Section 4 we look at the possibility if the CGB equation can be invariant
under arbitrary reparametrization of time at all.  Section 5 is devoted to
the case when the CGB equation is invariant under a Kac-Moody
algebra. Some
conclusions are presented in Section 6.
\section{Allowed transformations and a canonical generalized
Burgers equation} We shall use  "\at{}" (or equivalence transformations) to map \eqref{1.1} to some simple form. These
transformations are defined to be (invertible) point transformations
\begin{equation}\label{gat}
\tilde{x}=X(x,y,t,u),\quad \tilde{y}=Y(x,y,t,u),\quad \tilde{t}=T(x,y,t,u),\quad \tilde{u}=U(x,y,t,u),
\end{equation}
taking equations of
the form \eqref{1.1} into another equations of the same form, but
possibly with different coefficient functions.
That is, the
transformed equation  will  be the same as eq. \eqref{1.1}, but
the arbitrary functions can change. The typical features of
the equation are that the new functions $\tilde{p}(\tilde{t})$
and $\tilde{q}(\tilde{t})$ depend on $\tilde{t}$ alone, the
others on $\tilde{y}$ and $\tilde{t}$, but no $\tilde{x}$
dependence is introduced. The only $\tilde{t}$-derivative is
$\tilde{u}_{\tilde{x}\tilde{t}}$, the only nonlinear term is
$\tilde{p}(\tilde{t})(\tilde{u}\tilde{u}_{\tilde{x}})_{\tilde{x}}$
and the only derivative higher than a second order one is
$\tilde{q}(\tilde{t})\tilde{u}_{\tilde{x}\tilde{x}\tilde{x}}$.
These form-preserving conditions  restrict  \eqref{gat} to the form (the so-called local fiber-preserving transformations)
\begin{equation}\label{2.1}
\begin{array}{ll}
& u(x,y,t)=R(t)\tilde{u}(\tilde{x},\tilde{y},\tilde{t})-
\displaystyle\frac{\dot{\alpha}}{\alpha p}x+S(y,t),\\[.3cm]
& \tilde{x}=\alpha(t)x+\beta(y,t),\quad \tilde{y}=Y(y,t),\quad
\tilde{t}=T(t),\\[.3cm]
& \alpha\ne 0,\quad  R\ne 0\quad Y_{y}\ne 0,\quad \dot{T}\ne
0,\quad \dot{\alpha}f(y,t)=0.
\end{array}
\end{equation}
We note that the constraint $\dot{\alpha}f(y,t)=0$ should be imposed for the new $\tilde{h}$ to be $x$-independent. The new coefficients in the transformed equation satisfy
\begin{equation}\label{2.2}
\begin{aligned}
  \tilde p(\tilde t) &  = p(t)\frac{{R\alpha }}
{{\dot T}},\quad \tilde q(\tilde t) = q(t)\frac{{\alpha ^2 }}
{{\dot T}}, \\
  \tilde \sigma (y,t) &  = \sigma (y,t)\frac{{Y_y^2 }}
{{\alpha \dot T}}, \\
  \tilde a(\tilde y,\tilde t) &  = \frac{1}
{{\alpha \dot T}}\{ a Y_y  + \sigma Y_{yy} \} , \\
  \tilde b(\tilde y,\tilde t) &  = \frac{1}
{{\alpha \dot T}}\{ (b\alpha  + 2\sigma \beta _y )Y_y  + \alpha Y_t \} , \\
  \tilde c(\tilde y,\tilde t) &  = \frac{1}
{{\alpha \dot T}}\{ c\alpha ^2  + \beta_t   \alpha  + pS\alpha
^2  + \sigma \beta _y^2
+ b\alpha \beta _y \} , \\
  \tilde e(\tilde y,\tilde t) &  = \frac{1}
{{\alpha R\dot T}}\{ R\alpha e - R\dot \alpha  + \dot R\alpha  + aR\beta _y  + \sigma R\beta _{yy} \} , \\
  \tilde f(\tilde y,\tilde t) &  = \frac{1}
{{\alpha \dot T}}f, \\
  \tilde h(\tilde y,\tilde t) &  = \frac{1}
{{\alpha R\dot T}}\{ h - \frac{d}{dt}\left( {\frac{{\dot \alpha }}
{{\alpha p}}} \right)   + \frac{1}{p}\left( {\frac{{\dot \alpha }} {{\alpha
}}} \right)^2  + \sigma S_{yy}  + aS_y  + fS - e\frac{{\dot
\alpha }}{{\alpha p}}\}.  \\
\end{aligned}
\end{equation}

We now choose the functions $R(t), T(t)$ and $Y(y,t)$  in eq.
\eqref{2.1} to satisfy

\begin{equation}\label{2.4}
\begin{gathered}
  \dot T(t) = q(t)\alpha ^2 (t),\quad R(t) = \frac{q}
{p}\alpha ,  \hfill \\
  Y_y  = \alpha ^{3/2} \sqrt {\left| {\frac{{q(t)}}
{{\sigma (y,t)}}} \right|}  \hfill \\
\end{gathered}
\end{equation}
and thus normalize

\begin{equation}
\tilde p(\tilde t) = 1,\quad\tilde q(\tilde t) = 1,\quad\tilde
\sigma (\tilde y,\tilde t) = \varepsilon  =  \mp 1.
\end{equation}
By an appropriate choice of the functions $\beta(y,t)$ and
$S(y,t)$ we can arrange  to have
$$\tilde{e}(\tilde{y},\tilde{t})=\tilde{h}(\tilde{y},\tilde{t})=0.$$
Finally, equation \eqref{1.1} is reduced to its canonical form

\begin{equation}\label{canon}
\begin{split}
&(u_t+uu_x+u_{xx})_x+\varepsilon u_{yy}+a(y,t)u_y+b(y,t)u_{xy}\\
&+ c(y,t)u_{xx}+f(y,t)u=0,\quad  \varepsilon=\pm 1.
\end{split}
\end{equation}
With no loss of generality we can restrict our study to symmetries
of eq. \eqref{canon}. All results obtained for eq. \eqref{canon}
can be transformed into results for eq. \eqref{1.1}, using the
transformations \eqref{2.1}. We shall call eq. \eqref{canon} the
"canonical generalized Burgers equation" (CGB).

We mention that Lie point transformations are particular
cases of \at{}. When the form of the coefficients is
preserved, \at{} coincide with symmetry \tr s of the equation.

\section{Determining equations for the symmetries}
We restrict ourselves to Lie point symmetries. The Lie algebra
of the symmetry group is realized by vector fields of the form

\begin{equation}\label{3.1}
\hat{\mathbf{V}}=\xi\gen x+\eta\gen y+\tau\gen t+\phi\gen u,
\end{equation}
where $\xi$, $\eta$, $\tau$ and $\phi$ are functions of $x,y,t$
and $u$.
To determine the form of $\hat{\mathbf{V}}$ we apply the standard infinitesimal algorithm (see, for instance, Olver's book \cite{Olver91}) which basically consists of  requiring that the third prolongation ${\rm pr}^{(3)}\hat{\mathbf{V}}$ of the vector field on the third jet space  should annihilate the equation  on its solution manifold. This requirement provides an overdetermined set of linear partial differential equations for the coefficients $\xi$,
$\eta$, $\tau$ and $\phi$ in  \eqref{3.1}.

For eq. \eqref{canon} these equations  which  do not
involve the functions $a, b, c$ and $f$ can be solved and find that the general element of the
symmetry algebra has the form
\begin{equation}\label{3.2}
\hat{\mathbf{V}}=\tau(t)\gen t+(\frac{1}{2}\dot{\tau}x+\xi_0(y,t))\gen x +(\frac{3}{4}\dot{\tau}y+\eta_0(t))\gen y+
(-\frac{1}{2}\dot{\tau}u+\frac{1}{2}\ddot{\tau}x+S(y,t))\gen u,
\end{equation}
where
\begin{equation}\label{3.3}
S(y,t) =  - \tau c_t  - (\frac{3} {4}\dot \tau y + \eta _0 )c_y +
\xi _{0,t}  + b\xi _{0,y}  - \frac{1} {2}c\dot \tau.
\end{equation}
The remaining determining equations for $\tau(t)$, $\eta(t)$ and
$\xi_0(y,t)$ are
\begin{align}
  & {4\tau a_t  + (3\dot \tau y + 4\eta _0 )a_y  + 3a\dot \tau  = 0}, \hfill\label{3.4}
  \\[.3cm]
  & { - 4\dot \eta _0  - 3y\ddot \tau  + 4\tau b_t  + (3\dot \tau y + 4\eta _0 )b_y
  + b\dot \tau  - 8\varepsilon \xi _{0,y}  = 0}, \hfill \label{3.5}
  \\[.3cm]
  & {a\xi _{0,y}  + \varepsilon \xi _{0,yy}  = 0}, \hfill\label{3.6}
  \\[.3cm]
  & {f\ddot \tau  = 0}, \hfill \label{3.7} \\[.3cm]
  & {6f\dot \tau  + 4f_t \tau  + f_y (3\dot \tau y + 4\eta _0 ) = 0}, \hfill\label{3.8}
  \\[.3cm]
  & {2\dddot \tau  + 4fS + 4aS_y  + 4\varepsilon S_{yy}  = 0}.
  \hfill\label{3.9}
 \end{align}

At this juncture, there are different directions to go for dealing with determining equations. One is to perform a complete symmetry
analysis of eqs. \eqref{3.3},...,\eqref{3.9} for arbitrary (given)
functions $a$, $b$, $c$ and $f$. Of course, one can well proceed to determine the coefficients given that the equation is invariant under low-dimensional Lie algebras. Works in this direction for equations having dependence on several arbitrary functions of both independent and dependent variables and their derivatives exist in the literature (see for example \cite{Gungor04-1, Gungor04-2, Gungor08-3}). This approach requires the knowledge of structural results on the  classical Lie algebras.
Here we shall take another approach and determine the
conditions on these functions that permit the symmetry algebra to
be infinite-dimensional. This will happen when at least one of the
functions $\tau(t)$, $\eta_0(t)$ and $\xi_{0}(y,t)$ remains an
arbitrary function of at least one variable.

\section{Search for the Virasoro symmetries of the CGB equation}
We are looking for conditions on the coefficients $a,
b, c$ and $f$ that allow equations \eqref{3.4},...\eqref{3.9} to
be solved without imposing any conditions on $\tau(t)$. Below we shall see that this can not be realized for any possible choice of the coefficients.

From eq. \eqref{3.7} we see that $\tau$ is linear in $t$, unless
we have $f(y,t)\equiv 0$. Once this condition is imposed,
equations \eqref{3.7} and \eqref{3.8} are solved identically. Eq.
\eqref{3.4} leaves $\tau(t)$ free if either we have $a=0$, or
$a=a_0(y+\lambda(t))^{-1}$ where $a_0\ne 0$ is a constant and
$\lambda(t)$ is some function of $t$. We investigate the two
cases separately. First let us assume
\begin{equation}\label{4.1}
a=\frac{a_0}{y+\lambda(t)},\quad a_0\ne 0.
\end{equation}
Then we view eq. \eqref{3.4} as an equation for $\eta_0(t)$ and
obtain
\begin{equation}\label{4.2}
\eta_0(t)=\frac{1}{3}(2\lambda \dot{\tau}-3\dot{\lambda}\tau).
\end{equation}
From eq. \eqref{3.6} we see $\xi_0(y,t)$ may be an arbitrary function of
$t$, but never of $y$ (we have $\varepsilon=\pm 1$).  Three possibilities for $\xi_0(y,t)$ occur:

\noindent 1.) $a_0\varepsilon\ne \pm 1$

\begin{equation}\label{4.3}
\xi _0  = \frac{1}{1 - a_0 \varepsilon }  \mu _1 (t)(y + \lambda )^{ -
a_0 \varepsilon  + 1}   + \mu _0 (t).
\end{equation}

\noindent 2.) $a_0\varepsilon=1$

\begin{equation}\label{4.4}
\xi _0  =   \mu _1 (t)\ln (y + \lambda ) + \mu _0 (t).
\end{equation}

\noindent 3.) $a_0\varepsilon=-1$

\begin{equation}\label{4.5}
\xi _0  =   \mu _1 (t) (y + \lambda )^2  +
\mu _0 (t).
\end{equation}
We must now put $\xi_0$ of \eqref{4.3}, \eqref{4.4} or \eqref{4.5}
into eq. \eqref{3.5} and solve the obtained equation for
$\mu_1(t)$. The expression for $\mu_1(t)$ must be independent of
$y$ for all values of $\tau$. Moreover, for $\tau(t)$ to remain
free, there must be no relation between $b(y,t)$ and $\tau(t)$.
These conditions cannot be satisfied for any value of
$a_0\varepsilon$. Hence, if $a(y,t)$ is as in eq. \eqref{4.1} the
generalized Burgers equation \eqref{canon} does not allow a Virasoro
algebra.

The other case to consider is $a=0$ (in addition to $f=0$). Eq.
\eqref{3.6} is easily solved in this case and we obtain
\begin{equation}\label{4.6}
\xi _0(y,t)  =  \mu_1 (t)y + \mu _0 (t)
\end{equation}
with $\mu_1(t)$ and $\mu_0(t)$  arbitrary. We insert $\xi_0(y,t)$
into eq. \eqref{3.5} and try to solve for $\mu_1(t)$. This is possible
if and only if we have $b=b_1(t)y+b_0(t)$.
On the other hand,  the $y$ independent coefficient of \eqref{3.5} restricts the form of $\tau$ which implies that no Virasoro algebra can exist at all. In the following analysis we shall see that in that case there can exist at most two arbitrary
functions.

\begin{thm}\label{t1}
The canonical \gbe{} \eqref{canon} can never allow the Virasoro algebra as
a symmetry algebra for any choice of the coefficients.
\end{thm}

\section{Kac-Moody symmetries of the CGB e\-qua\-tion}
In section 4 we have shown that  the symmetry algebra of the
canonical generalized Burgers equation cannot  contain a Virasoro
algebra. In this section
we will determine the conditions on the functions $a(y,t)$,
$b(y,t)$, $c(y,t)$ and $f(y,t)$ under which the CGB equation only
allows a Kac-Moody algebra. Thus,
the function $\tau(t)$ will not be free, but $\eta_0(t)$ of eq.
\eqref{3.2} will be free, or $\xi_0(y,t)$ will involve at least one
free function of $t$.
\subsection{The function $\eta_0(t)$ free}
Eq. \eqref{3.4} will relate $\eta_0$ and $a(y,t)$ unless we have
$a_y=0$. Hence we put $a_y=0$. For $a=a(t)\ne 0$ eq. \eqref{3.4}
implies $\tau(t)=\tau_0a^{-4/3}$. Eq. \eqref{3.6}  yields
$$\xi_0(y,t)=\xi_1(t)e^{-a\varepsilon y}+\xi_0(t).$$
Eq. \eqref{3.5} then provides a relation between $\eta_0(t)$ and
$b(y,t)$. Hence $\eta_0(t)$ is not free. Thus, if $\eta_0(t)$ is to
be a free function, we must have $a(y,t)=0$. Eq. \eqref{3.4} is
satisfied identically. From eq. \eqref{3.6} we have
\begin{equation}\label{5.1}
\xi _0 (y,t) =   \rho(t)y + \sigma (t).
\end{equation}
Eq. \eqref{3.5} will leave $\eta_0$ free only if we have
\begin{align}
b(y,t)&=b_1(t)y+b_0(t),\label{5.2}\\[.2cm]
\rho(t)&=\frac{\varepsilon}{8}(-4\dot{\eta_0}+4\tau \dot{b}_0+4\eta_0
b_1+b_0 \dot{\tau}),\label{5.3}\\[.2cm]
3\ddot{\tau}-4{(\tau b_1)}^{.}&=0\label{5.4}.
\end{align}
For $f\ne 0$ we have $\ddot{\tau}=0$ and eq. \eqref{3.9} will
relate $\eta_0(t)$ to $c(y,t)$, $b_1$ and $b_0$. Thus, for
$\eta_0(t)$ to be free, we must have $f(y,t)=0$. Eq. \eqref{3.9}
reduces to
\[
-2\varepsilon\dddot \tau  + \dot{\tau}(8c_{yy}+3yc_{yyy})+4(\tau c_{yyt}+\eta_0 c_{yyy})  = 0.
\]
$\eta_0(t)$ is free if we have
\begin{align}
& c(y,t) = c_2 (t)y^2  + c_1 (t)y + c_0 (t), \label{5.5}\\[.3cm]
& -2\varepsilon \dddot \tau  + (8\tau \dot{c_2} + 16\dot \tau c_2)
= 0.\label{5.6}
\end{align}
The only equation that remains to be solved is eq. \eqref{5.6}.
Both functions $\eta_0(t)$ and $\sigma(t)$ remain free.  The
most general CGB equation allowing $\eta_0(t)$ to be a free
function is obtained if eq. \eqref{5.6} is solved identically by
putting $\tau=0$. Then $\eta_0(t)$ and $\sigma(t)$ are arbitrary. On the other hand, from \eqref{5.6} we see that $\tau(t)$ can not remain free. This again implies that  the symmetry algebra can by no means  be Virasoro  type.
Using eq. \eqref{3.2} and the above results with the identification $\eta=\eta_0$, $\xi=\sigma$ we obtain the
following theorem.
\begin{thm}\label{t4}
The equation
\begin{equation}\label{5.7}
\begin{split}
(u_t&+uu_x+u_{xx})_x+\varepsilon u_{yy}+(b_1(t)y+b_0(t))u_{xy}\\
&+(c_2(t)y^2+c_1(t)y+c_0(t))u_{xx}=0,
\end{split}
\end{equation}
where $\varepsilon=\pm 1$ and $b_0, b_1, c_0, c_1, c_2$ are
arbitrary functions of $t$, is the most general canonical \gbe{},
invariant under an infinite-dimensional Lie point symmetry group
depending on two arbitrary functions. Its Lie algebra has a
Kac-Moody structure  and is realized by vector
fields of the form

\begin{equation}\label{5.8}
\hat{\mathbf{V}}=X(\xi)+Y(\eta),
\end{equation}
where $\xi(t)$ and $\eta(t)$ are arbitrary smooth  functions of
time and
\begin{align}
X(\xi)&=\xi\gen x+\dot{\xi}\gen u, \label{5.9}\\
\begin{split}\label{5.10}
Y(\eta)&=\eta\gen
y+\frac{\varepsilon}{2}y(-\dot{\eta}+b_1\eta)\gen
x+\curl{[-2c_2\eta\\
&+\frac{\varepsilon}{2}(-\ddot{\eta}+\dot{b}_1\eta+b_1^2\eta)]y
-c_1\eta+\frac{\varepsilon}{2}b_0(-\dot{\eta}+b_1\eta)}\gen u.
\end{split}
\end{align}
\end{thm}
Several comments are in order:

\begin{enumerate}
\item Surprisingly enough, the Kac-Moody symmetry algebras  of the canonical generalized KP Eq. \eqref{GKP}-\eqref{SCoef} with $b, c$ given by \eqref{KMT} and Burgers (Eq. \eqref{5.7}) equations coincide.
\item As in the GKP case, equation \eqref{5.7} can be further simplified by
\at{}. Indeed, let us restrict the transformation \eqref{2.1} to
\begin{equation}\label{5.11}
\begin{gathered}
  u(x,y,t) = \tilde u(\tilde x,\tilde y,\tilde t) + S_1 (t)y + S_0 (t), \hfill \\
  \tilde x = x + \beta _1 (t)y + \beta _0(t) ,\;\;\tilde y = y + \gamma (t),\;\;\tilde t = t. \hfill \\
\end{gathered}
\end{equation}
For any functions $b_1(t)$ and $c_2(t)$ we can choose $S_1, S_0,
\beta_0, \beta_1$ and $\gamma$ to set $b_0, c_1$ and $c_0$ equal
to zero. Thus, with no loss of generality, we can set
\begin{equation}\label{5.12}
b_0(t)=c_1(t)=c_0(t)=0
\end{equation}
in eq. \eqref{5.7}, \eqref{5.9} and \eqref{5.10}.

\item Let us now consider the cases when eq. \eqref{5.7} has an additional symmetry. To do this we should solve
equations \eqref{5.4}-\eqref{5.6}.

{\bf Case 1.} $b_1=0, c_2\ne 0$

We assume \eqref{5.12} is already satisfied. From  \eqref{5.4}
we have $\tau=\tau_1 t+\tau_0$ and from \eqref{5.6}
$$ c_2=k \tau^{-2}=k(\tau_1t+\tau_0)^{-2}$$ where $\tau_1$, $\tau_0$, $k$ are constants.  The additional symmetry is
\begin{equation}\label{5.13}
T=(\tau_1t+\tau_0)\gen t+\frac{1}{2}\tau_1 x \gen x+\frac{3}{4}\tau_1 y\gen y-\frac{1}{2}\tau_1u\gen u.
\end{equation}
Under translation of $t$, it is equivalent to the dilatational  symmetry
$$D=t\gen t+\frac{1}{2} x \gen x+\frac{3}{4} y\gen y-\frac{1}{2}u\gen u.$$

{\bf Case 2.} $b_1\ne 0$, $b_0=c_1=c_0=0$

Eq. \eqref{5.4} can be integrated to give a first order linear equation for $\tau$
in terms of $b_1$ and eq. \eqref{5.6}
provides the constraint between $b_1$ and $c_2$
\begin{equation}\label{5.14}
\frac{d}{dt}(\tau^2 c_2)=\frac{\varepsilon}{4} \tau \frac{d^2}{dt^2}(\tau b_1).
\end{equation}
The additional element of the symmetry
algebra in this case is
\begin{equation}\label{5.15}
T=\tau\gen t+\frac{1}{2}\dot{\tau}x \gen x+\frac{3}{4}\dot{\tau}y\gen y+[\frac{1}{2}\ddot{\tau}x-(\tau \dot{c_2}+2c_2 \dot{\tau})y^2-\frac{1}{2}\dot{\tau}u]\gen u
\end{equation}
with $\tau$ being a solution of
$$\dot{\tau}-\frac{4}{3}b_1\tau=k.$$
\end{enumerate}

\subsection{One free function in symmetry algebra}
We have established that if $\tau(t)$ is free in eq. \eqref{3.2},
then there are three free functions. If $\tau$  is not free, but
$\eta_0(t)$ is, then there are two free functions. Now let
$\tau(t)$ and $\eta_0(t)$ be constrained by the determining
equations, but let some freedom remain in the function
$\xi_0(y,t)$.

First of all we note that if we put

\begin{equation}\label{5.16}
\tau=0, \quad \eta_0=0,\quad \xi_0(y,t)=\xi(t)
\end{equation}
in eq. \eqref{3.2} then eqs. \eqref{3.4},...,\eqref{3.8} are
satisfied identically and eq. \eqref{3.9} reduces to

\begin{equation}\label{5.17}
f\dot{\xi}=0.
\end{equation}
Hence

\begin{equation}\label{5.18}
X(\xi)=\xi(t)\gen x+\dot{\xi}(t)\gen u,
\end{equation}
with $\xi(t)$ arbitrary, generates Lie point symmetries of the
CGB equation for $f(y,t)=0$ and any functions $a(y,t), b(y,t)$,
and $c(y,t)$.

For $f\ne 0$ we have $\tau=\tau_1 t+\tau_0$ from eq. \eqref{3.7}.
Eq. \eqref{3.6} then determines the $y$ dependence of $\xi_0$.

We skip the details here and just state that the remaining
equations \eqref{3.5}, \eqref{3.8} and \eqref{3.9} do not allow
any solutions with free functions.

We state this result as a theorem.
\begin{thm}\label{t5}
The CGB equation \eqref{canon} is invariant under an
infinite-di\-men\-sion\-al Abelian group generated by the vector
field \eqref{5.18} for $f(y,t)=0$ and $a, b, c$ arbitrary.
\end{thm}

Theorems  \ref{t4} and \ref{t5} sum up all cases when the
symmetry algebra of the CGB equation is infinite-dimensional.
\section{Applications and conclusions}
We have identified all cases when the generalized Burgers equation has
an infinite-dimensional symmetry group. Let us now discuss the
implications of this result.

\subsection{Equation with nonabelian Kac-Moody symmetry algebra}
The symmetry algebra \eqref{5.8} of eq. \eqref{5.7} is
infinite-dimensional and nonabelian. Indeed, we have
\begin{equation}\label{6.2}
[Y(\eta_1),Y(\eta_2)]=X(\xi),\quad
\xi=-\frac{\varepsilon}{2}(\eta_1\dot{\eta}_2-\dot{\eta}_1\eta_2).
\end{equation}
We can  apply the method of symmetry reduction
to obtain particular solutions. The operator $X(\xi)$ of eq.
\eqref{5.9} generates the transformations
\begin{equation}\label{6.3}
\tilde{x}=x+\lambda \xi(t),\quad \tilde{y}=y,\quad
\tilde{t}=t,\quad
\tilde{u}(\tilde{x},\tilde{y},\tilde{t})=u(x,y,t)+\lambda
\dot{\xi}(t),
\end{equation}
where $\lambda$ is a group parameter. We see that \eqref{6.3} is
a transformation to a frame moving with an arbitrary acceleration
in the $x$ direction. For $\xi$ constant this is a translation,
for $\xi$ linear in $t$ this is a Galilei transformation. An
invariant solution will have the form
\begin{equation}\label{6.4}
u=\frac{\dot{\xi}}{\xi}x+F(y,t).
\end{equation}
Substituting into eq. \eqref{5.7} we obtain the family of
solutions
\begin{equation}\label{6.5}
u=\frac{\dot{\xi}}{\xi}x-\frac{\varepsilon}{2}
\;\frac{\ddot{\xi}}{\xi} y^2+ \rho(t)y+\sigma(t)
\end{equation}
with $\rho(t)$ and $\sigma(t)$ arbitrary.

The transformation corresponding to the general element
$Y(\eta)+X(\xi)$ with $\eta\ne 0$ is easy to obtain, but more
difficult to interpret. An invariant solution will have the form
\begin{equation}\label{6.6}
\begin{split}
u&=[-c+\frac{\varepsilon}{4}(\dot{b}+b^2-\frac{\ddot{\eta}}{\eta})]y^2
+\frac{\dot{\xi}}{\eta}y+F(z,t)\\
z&=x+\frac{\varepsilon}{4}(-b+\frac{\dot{\eta}}{\eta})y^2-\frac{\xi}{\eta}y.
\end{split}
\end{equation}
We have put $b_1=b,\; c_2=c,\; b_0=c_1=c_0=0$, which can be done with
no loss of generality. We now put $u$ of eq. \eqref{6.6} into eq.
\eqref{5.7} (for $c_1=c_0=b_0=0$) and obtain the reduced equation
\begin{equation}\label{6.7}
(F_t+F F_z+F_{zz})_z+\varepsilon
\frac{\xi^2}{\eta^2}F_{zz}+\frac{1}{2}(\frac{\dot{\eta}}{\eta}-b)F_z
-2c\varepsilon+\frac{1}{2}(\dot{b}+b^2-\frac{\ddot{\eta}}{\eta})=0.
\end{equation}
Putting
\begin{equation}\label{6.8}
\begin{aligned}
  F(z,t) &  = \tilde F(\tilde z,\tilde t),\;\;\;\tilde z = z + \beta (t),
  \;\;\;\tilde t = t, \\
  \dot \beta (t) &  =  - \varepsilon \frac{{\xi ^2 }}
{{\eta ^2 }} \\
\end{aligned}
\end{equation}
we eliminate the $F_{zz}$ term. Choosing $\dot{\eta}/\eta=b(t)$ we
obtain the equation
\begin{equation}\label{6.9}
(F_t+F F_z+F_{zz})_z=2\varepsilon c(t),
\end{equation}
an equation that is not integrable (for $c\ne 0$)  We note
that for $c=0$ \eqref{6.9} reduces to the 1-dimensional Burgers equation.

\subsection{Comments}
By the results of this paper we have shown that neither 2+1-dimensional Burgers equation nor its generalizations of the form \eqref{1.1} can allow a Virasoro type symmetry group. The largest infinite-dimensional symmetry  allowed can be Kac-Moody type. In addition to this, for specific choice of the coefficients it has one more symmetry.   It should also be worthwhile reiterating the overlapping of the  Kac-Moody symmetries allowed by generalized KP and Burgers equations.

The most ubiquitous symmetry of the generalized Burgers equation is the
transformation \eqref{6.3} to an arbitrary frame moving in the $x$
direction. Its presence only requires the coefficient $f(y,t)$ in
eq. \eqref{1.1} for $p=1$ or in \eqref{canon} to be $f(y,t)\equiv 0$.
Invariance of a solution under such a general transformation is
very restrictive and leads to solutions that are at most linear in
the variable $x$ and have a prescribed $y$ dependence (see
solutions \eqref{6.5}).

The transformations generated by $Y(\eta)$ leave a more restricted
class of generalized Burgers equations invariant, those of eq. \eqref{5.7}. The
invariant solutions have the form \eqref{6.6}.   They are obtained by solving the reduced equation eq. \eqref{6.7} with $F_{zz}$ transformed away or \eqref{6.9}. For general $c(t)$, this is
difficult, but for $c(t)=0$ this is just the Burgers equation, for
arbitrary $b(t)$, as long as we choose $\dot{\eta}/\eta=b(t)$. Any
solution of the Burgers equation will, via eq. \eqref{6.6}, provide $y$
dependent solutions of the corresponding generalized Burgers equation.

One-dimensional additional subalgebras can be imbedded   into Kac-Moody subalgebras to form two-dimensional subalgebras. Invariance under them will lead to reductions to ODEs (see \cite{Gungor01-2}).


\end{document}